# METHODS AND SYSTEMATIC REFLECTIONS

## The Fundamental Principles of Existence and the Origin of Physical Laws

Grandpierre, Attila, *Konkoly Observatory, Hungarian Academy of Sciences, Budapest, HUNGARY*
H-1525 Budapest, P. O. Box 67.
Grandp@konkoly.hu

**ABSTRACT.** In this essay the ontological structure of reality is explored. The question of reducibility of biology to physics is considered in the context of their ultimate principles. It is shown that biology is an ontologically autonomous science and is based on its own, independent ultimate principle that is independent from that of physics. In the next step it is shown that self-consciousness represents a separate realm with its own, ontologically autonomous, ultimate principle. Understanding that reality is based on ultimate principles, a new possibility arises to interpret the origin of physical laws.

1. **INTRODUCTION**

Our concept of the universe and the material world is foundational for our thinking and our moral lives.  In an earlier contribution to the URAM project I presented what I called 'the ultimate organizational principle' of the universe.  In that article (Grandpierre 2000, pp. 12-35) I took as an adversary the wide-spread system of thinking which I called 'materialism'.  According to those who espouse this way of thinking, the universe consists of inanimate units or sets of material such as atoms or elementary particles.  Against this point of view on reality, I argued that it is 'logic', which exists in our inner world as a function of our mind, that is the universal organizing power of the universe.  The present contribution builds upon this insight.  Then I focussed on rationality; now I am interested in the responsibility that is the driving force behind our effort to find coherence and ultimate perspectives in our cosmos.

Our existence in the present consumer societies requires a way to achieve coherence between our lifelong activity of discrete deeds and the overall meaning of our whole life.  I suggest here two ultimate positions on responsibility, which I will call the Alpha and Omega Viewpoints.

1.1 *The Alpha and Omega Viewpoints*

The Alpha Viewpoint starts from the point from where our responsibility emerges. First of all, we are responsible for our lives, to fulfil their genuine meaning. Moreover, we are also responsible for our family, for our natural collectives (friends, circles, and communities), for our nation, society, and for humankind. Human rights are one important element here, but the Alpha Viewpoint goes beyond them. For instance, we are responsible for our environment, for the biological and spiritual meaning present in nature and the universe. Our lives are connected to the bio-spiritual activity of nature and the universe. We are born to realize a life that should fulfil the spiritual task of the cosmic organizational principle creating our personality through constructing the special combination of our parents' genes that led to our development and birth into this world. This is the viewpoint of personal responsibility, which I call the Alpha Viewpoint of Ultimate Reality and Meaning.

The Omega Viewpoint has a different perspective. It recognizes that we are not only personal, individual beings. Our personality is possible only through the mediation of some more common traits like lingual-cultural-social-national identity. At the same time, we are actually representatives of *Homo Sapiens Sapiens*, unelected delegates of the living kingdom, part and parcel of nature and of the cosmos. We cannot be persons outside the world - inevitably we are part of a more comprehensive collectivity, namely, the cosmic, natural, social communities and families. Therefore we are presented with invaluable gifts from these natural configurations. Nature acts from within ourselves through the gifts of our genes, through our inner drives to live a meaningful life, to re-connect with the world considered as the other half of our personality. Therefore, on the basis of moral mutuality, we are responsible for the well-being of our natural drives towards completion and meaning; we are obliged to be active and find creative answers, to give back the gift of our lives to the nation, mankind, nature and the cosmos. Only by filling our lives with meaning in relation to these gifted natural communities can we become morally responsible partners to the life-giving and mind-giving factors of our lives.

Moreover, our lives in themselves have only a relative significance. They achieve their true significance only in relation to Ultimate Reality. There is, for instance, a difference for nature and the cosmos whether mankind has or has not existed. For this and other reasons we cannot let civilization drift in just any direction. We are obliged to involve ourselves in the generation of a civilization coherent with the genuine destination of mankind, and consistent with the authentic life giving and mind-enlivening factors of nature and the universe. I call this viewpoint as Omega Viewpoint - the decisive viewpoint of our lives, which offers the ultimate perspective in determining our activity of our human lives.

1.2 *Materialism*

According to the people whom I choose to describe as materialists, objective processes determine our actions and therefore we cannot be responsible for how we live them out. I regard materialism as anti-human warfare against the life we should live in accord with our deepest nature; it is a struggle against our morality and responsibility for our future, for the future of our family, of our collective, nations, mankind, nature and universe.

I offer an explicit presentation of the basic beliefs of physicalism and confront it with three ultimate principles: they are the principles of existence, life and self-consciousness. They are not new values; in them modern science and ancient Greek and pre-Greek science seem to meet. The ultimate principle of life is shown to be independent from the ultimate principle of physics. The question of the behaviour of matter, as directed by the physical principle, leads us to the possible explanation of the origin of physical laws.

# 2. ON THE PRECONDITIONS OF PRESENT DAY SCIENCE

The preconditions of science require a special kind of realism, one that is based on a totalitarian dehumanisation (see Bunge, 1967, 291ff). But is it scientific to require an objective world, the objectivity of which is based on a total independence from all spheres of human nature, and which allows the world only a material activity and essence? Could it be scientific to state that nothing is real in *Homo sapiens* except its inanimate matter? Could it be scientific to claim that if any other reality is occasionally present in humans, it is not present in the scientifically 'real' world?

Materialistic science in face of the real world has three basic constituents: corporeal building blocks, the 'least action' principle, and chance as the blind organisational factor. Where can we find a place for man in this materialistic world? Would it be in the realm of inert material objects? Or could (s)he find a role to play in realising the 'least action'? Or does (wo)man realize blind chance? Do we recognize man in these realms or in their combination?

Also known as the Maupertuis or Hamilton principle, 'least action' is the ultimate rule of physics. Modern physics formulates it as the 'variational principle' or "action principle", and from it all the laws of motion and energy are derived (Landau & Lifschitz 1959, Vol I, p. 12). 'According to this principle the objects studied by physics strictly follow the principle of least action. The implication of this is that a "materialistic natural philosophy" can be built based on physics according to which an inanimate world can be constructed in which matter is completely inert, inactive and unable to express any self-directed activity' (Grandpierre 2000b, p.15). The total physical equilibrium for a living being is the state of death. But the principle can act only if something moves the corporeal objects away from equilibrium. To make this possible, materialism needs a chance process. According to materialistic theorists, chance being blind (isotropic) can bridge any gap.

How should life develop from this material world? The implicit assumption of materialism is that this takes place by chance. The chance character of atomic collision is suggested by the kinetic gas theory, by the chaotic character of Brownian motion, and 'the sophisticated experiments of Aspect and Grangier (1986, 1987) involving the seemingly random directing of photons that impinge on a beam-splitter' (Quay 1995, p.10). Nevertheless, until now, materialistic science has not studied how the nature of chance is modified when the conditions for life are present. And since the blindness of the chance is a requisite, some scientists declare that blind chance has led to the development of life (Atkinson 1981). The factual presence of life argues for the presence of systematic chance as a substantial difference from blind chance.

For instance, materialistic science can allow for the possibility that a chair might be raised through the random character of the heat motion of its atoms. Nevertheless, the lifting up of a chair by just one millimetre would be possible with such a small probability that it is, in fact, ruled out of the range of real phenomena. If it is unlikely that a chair might rise from the ground thanks to its heat, then the generation of a living organism would require a much more improbable 'spontaneous' series of events (Hoyle, Wickramasinghe 1999a, b). If the 'spontaneous levitation' of a chair through quantum fluctuations should be regarded as an extremely improbable event, we could call such an event a 'miracle'. Since the events necessary for the development of life are each in themselves extremely improbable, they may each be regarded as wonders. In point of fact, materialistic science does not concern itself with marvels. However, the phenomenon of life does require the co-operation of 'miracles' that are even more improbable, and this co-operation occurs with a remarkably high degree of freedom, capable of organising a cosmic amount of information by its activity. The contingency involved in life on planet Earth would seem to put paid to the idea of blind chance being totally responsible for the material universe.

Moreover, these co-operative 'miracles' not only come into being but are also maintained and become organised systematically in a practical way throughout the life of an organism. The organism is governed by integrative principles that are completely different from the physical principle of the 'least action'. If we consider that the development of such wonders would need a significant amount of time, and that during this period nothing can 'switch off' the physical principle, we have to realize that if it were only a material principle directing the behaviour of matter, miracles would never happen. If they were to do so by random perturbations, then they could not co-operate with each other independently from the laws of the material principle attracting them back immediately towards the physical equilibrium of the state of death. The co-operating and ordered coherency in the activity of the allegedly random perturbations should be a highly transient and extremely rare occasion, in contradiction with the systematic and lifelong organizational activity of the organism against the approach towards physical equilibrium. Therefore, when we confront the material principle with the phenomenon of life, we realize that the principle of life is independent from the material principle of 'least actions'.

## 2.1 *The hidden natural philosophy*

Contemporary science - apparently without any scientific reason - has given up its earlier program of comprehensively understanding Nature. For the ancient Greeks, a comprehensive knowledge of Nature was a central feature of their philosophy; today's fundamental research is more modest, and predominantly materialistic in outlook. Contemporary physicists are mistaken when they narrow science. They are in error when they argue that because all entities of nature have physical attributes, therefore all natural entities are exclusively or essentially of a physical nature. This is a factual as well as a philosophical mistake.

Let us begin with the factual error. How far has materialism got in answering such basic problems as the origins of life and consciousness? It has gone no further than presuming – without any scientific investigation whatsoever – that life and consciousness originated from the available material system by the assistance of purely accidental material processes (Atkinson 1981). Give the world its original corporeality or materiality, and then, thanks to 'chance', all of a sudden living matter appears. This is unlikely, and probably wrong. It has been proven (Hoyle, Wickremasinghe 1999 a, b) that for the accidental appearance of even the simplest amino acid, $10^{2000}$ more years would be needed than the whole age of the Universe. The *a fortiori* argument would go like this: if the accidental character of life is probably unable to create even the simplest amino acid, then how could it be responsible for more complex forms of life itself? Recently, Delsemme (1998) argued against Hoyle's hypothesis by introducing complex hypercycles, the dynamics of which could help generate amino acids. Unfortunately, this argument postulates such complex hypercycles that already should contain information assumed to be present only as their end product. Therefore Delsemme's line of reasoning fails to offer an explanation for the origin of life.

There is an even more fundamental problem with claiming that 'chance' is the most fundamental factor in nature. It is this: physics is not the optional terrain of accidental processes. The physical laws of motion assign a definite course to physical processes, which are expressed, for example, in the Second Law of Thermodynamics: closed physical systems are directed towards a total thermal balance. If we pour hot water into cold water, we get lukewarm water. Eventually, says the Second Law, all differences will disappear, and the whole universe will come to the terminal state of 'heat death'. Closed physical systems usually advance directly towards physical equilibrium. Throw a stone in the air, and it will fall down, because the forces affecting it drive it towards the state of equilibrium. Throw a bird in the air, and, as long as it is alive, it will not fall down but will fly freely in the air far away from the physical equilibrium. Physical equilibrium is valid in the case of living systems only when they have ceased to be living systems. For living systems, the motions prescribed by physical equations always lead towards death. Living systems are living precisely because their behaviour is not exclusively directed by physical forces, but by other forces which modify the internal physical conditions of the living system in such a way that it will be moved as far away as possible from physical equilibrium, for the longest possible time. Stephen Hawking's *A Brief History of Time* expresses the brutal and dangerously antihuman materialistic view that human beings are mere material objects whose behaviour will be exactly calculated by the much-anticipated Grand Unified Theory of physics. 'Yet if there really is a complete unified theory, it would also presumably determine our actions' (Hawking 1988, p. 13). In this materialistic concept the predominant direction is that of physical equilibrium, i.e. death. These people are convinced that all processes advance inevitably towards the unavoidable 'equilibrium death' of physics.

## **2.2** *The philosophical foundation of materialism*

For the past century and a half materialism has dominated scientific thinking. Therefore, it is relevant to ask whether or not the basic propositions of materialism have been proven, and to identify what these basic propositions might be. To realize this task, first we have to consider the notion of matter and determine its content in present day materialism. To recognize what is modern about contemporary materialism we should return briefly to ancient Greece.

The notion of matter for the early Greek materialists was substantially different from today's concept of matter. The early Greeks thought of matter as something that was alive, perceiving it to be a living, spontaneously active entity that operated independently. This concept was called *hylozoism*, a word derived from the Greek words *hylē* meaning 'wood, or matter' and *zoē* or 'life'. *Hylozoism* was central even in such later thinkers as Averroes, Bernardino Telesio, Paracelsus, Cardanus, Gassendi, and Giordano Bruno. The first Greek philosophers, the Milesian Thales (624-546 BC), Anaximandros (610-547 BC), and Anaximenes (588-524 BC) all taught that matter is alive.

This is close to shamanism, which is of much interest to contemporary anthropologists. It was said about Thales 'his doctrine was that the world is animate' (Diogenes Laertius, 1959, 27). If this doctrine were his and his only, then it would mean that it was not a widespread view among the early Greeks. However, Thales may have learned it from his Phoenician ancestors. Anaximenes expressed his hylozoism in the following way: 'as our soul - which is air - governs us, in such the same way the universe is governed by breathing and air'. Xenophanes (582-485? B.C.) described the whole Cosmos as moved by thought: 'Without any tiredness he shakes everything with his thoughts'. The significant Greek philosophical school of the Stoics returned to a kind of *hylozoism* which we could term today as *organicism*.

It would be easy to dismiss *hylozoism* a priori as a scientifically and philosophically inadequate theory. But contemporary western thought has failed to examine critically the view of Descartes and his disciples regarding the inert nature of matter. These statements, especially the one about the lifeless and inert nature of matter, have been accepted by many scientists virtually unexamined, although according to the general requirements of science, every statement should have been thoroughly tested and scientifically checked, and only after a positive conclusion to these inquiries should they be received. The experimental and theoretical examination of the nature of matter being inert or not, however, has not occurred up to now. Despite this fact, the principle of inert matter has, in an unfounded and so unscientific way, come to prevail. It is the basis of modern physics' claim to autocracy and exclusive dominance in the material world – including the world of the (seemingly?) living as well. Remarkably, it has become the basis of our contemporary physical world concept, according to which even biology can be deduced from physics.

Let us consider whether or not animals are really lifeless? And are humans? Let us also think about the empirical fact that the life principle is deeper than the physical one, because living beings are guided not by the physical, but by the biological life principle. We experience in the overwhelming majority of cases that living beings behave not as physical bodies but as self-initiating beings. Now, if the life principle is deeper than the physical principle, then can we say that matter is at all times unresponsive and inert? It has to be seen that the contemporary physical world concept wants to force upon us the idea of inert matter, and that there are essential arguments for the invalidity of this physical world concept. Regarding the evidence of the dominance of the biological principle over the physical one in living organisms, we have to admit that there do exist living beings of which being alive is essential to their nature and being animate offers a way out of physical inertness and inanimateness.

However, with the development of modern science the notion of matter underwent a fundamental change. The 'objective' concept of perfectly lifeless, inert matter became generally accepted as a huge and uncontrolled extrapolation of the properties of matter shown within the strictly restricted conditions of physical experiments on every existent without any serious investigation. This extrapolation may well need further thorough study. Let us now examine what science today means by 'matter' or 'life', and then try to draw the conclusions regarding the material nature of life.

Materialism does not define the concept of matter clearly and unambiguously. On the one hand, the atoms and the elementary particles constituting it are regarded as substance, and, on the other hand, the gravitational fields, the electro-magnetic fields, the weak and the strong fields are also thought of as substantial and significant entities in contemporary science. It seems that if science discovers any new entity, that also will be co-opted as material, and therefore equally inert and inanimate. But then, are the words 'material entities' and 'existents' merely synonyms? But, in point of fact there are entities that are not material. Life as an organizational power and the phenomena of consciousness, intention and intellectual consideration, are real existents even though they are not material entities in themselves. Materialism states that phenomena of life and consciousness are mere epiphenomena of the material substrates. But what is the essence of materiality? For one thing, it is clear that physics is different from biology in that physics examines the inanimate world while biology deals with the phenomena of the living kingdom. Thus, since our contemporary world apparently accepts physics as the ground of its 'scientific world view', the notion of matter is defined on the basis of physics; therefore matter is defined as a lifeless entity. This way the central proposition of materialism will be that it is only the phenomena of the inanimate world that are of any relevance in the understanding of the behaviour of all entities. This means that according to this materialistic concept, there can be only a quantitative and/or qualitative difference between the phenomena of life and those of the lifeless world, but not a substantial difference. The only reason for this qualitative difference would be the complex and specially arranged structures of living beings. Considering all this, we have to decide whether the difference between the phenomena of life and those of the inanimate world has to be regarded as substantial or not. To be able to answer this question, we have to uncover the concept of matter and what is its essence.

There is a quantitative difference between, for example, a stone and a heap of stones. The difference between a brick and a house is a qualitative difference, and this may involve an essential agreement as well as an essential difference. The essence is to be grasped in the nature of the elements of the house. If we are to trace it back to the nature of the elements of the house, then the house is nothing else but a heap of bricks. But if the essence is considered from the point of view of organisation, of function, of usability, then the difference is essential, for a brick is only a building block, while a house is a building. For us, one cannot live in a brick at all, whereas a house can usually be used for living in it. If, when examining the nature of living beings, their essence were given by the nature of their constituent parts, then living beings would be lifeless beings, for surely they are mostly (or totally) made up of inanimate (or apparently inanimate) atoms. The point of debate in this question is precisely whether living beings have any such plans, designs, intentions, such as atoms do not have at all, at least according to materialism. Thus the directionless atoms would have to develop into living beings with a plan as a result of chance, due to an extraordinary series of pure accidents. But can a plan appear just like that, blindly, out of perfectly unsystematic chance?

Although it seems that theoretically this cannot be deemed impossible, as there are some everyday instances to prove that seeming regularities arise from accidental occurrences, still this question should be examined more deeply and in more detail. What we are talking about here is not just throwing coins in the air and from time to time noticing regularities in the sequence of heads or tails coming down. Physical laws govern the phenomena of physics. The world of physical phenomena involves such a 'centre of attraction' in which any deviation from the state of physical equilibrium elicits the occurrence of a process aiming at the re-establishment of the state of equilibrium. And this equilibrium is perfectly lifeless, corresponds to a stone rolled down a hill, unavoidably attracted by the terminus of arriving at the lowest place at the bottom of the valley. Moreover, if by chance fluctuations were to occur, such as an upward jump, these could only be temporary and transitory, not affecting the gross overcome of the process, since sooner or later the stone would take a downward course. The fate of the rolling stone is determined, and the life of a rolling stone is formed by the attraction of the final equilibrium. All other phenomena can only be transitory and temporary, their time-scales being (much) smaller than that of the whole movement of the stone. Although a 'miracle' might occur so that a chair might rise from the ground thanks to the accidental collisions of atoms, still this would be an extremely improbable event, and the more improbable such an event is, the shorter will be the time that it will exist. The laws of physics themselves make sure that the improbability is extreme. The centre of attraction of the physical laws is well known - it is the state of equilibrium. If a living being as a system reaches the state of perfect and complete physical equilibrium, then it has already died – so the equilibrium state as a state of attraction in physics is, for us living beings, the state of death.

## 3. ESSENTIAL NOVELTY - FROM PHYSICS TO BIOLOGY

However extensive an area we assign to pure chance, we must point out that both the improbability of occurrence and the probability of maintenance of such a presumed state increase in proportion to that state's distance from equilibrium. The science of physics has no branch dealing with the physics of phenomena arising from extremely improbable events. To imagine having a branch of physics dealing with phenomena that are extremely improbable on a physical basis is to imagine a branch of science occupied with, for example, examining how sugar dissolved in tea could re-assemble into cubes of sugar. As a matter of fact, such a branch of physics is nonsense, since there are no laws of such extremely improbable phenomena. Therefore, the fact that there exist characteristic symptoms, recurrent lawful phenomena and explicit laws in the field of the phenomena of life, contradicts the universal validity of the basic principle of inanimateness. I suggest that we can talk about an essential novelty, a substantial novelty not deducible from the laws of the other level if on a new level the behaviour of a system becomes systematically different, and – while still observing the validity of the physical laws it has followed so far – the system now forms its behaviour following other laws instead of the physical laws.

Biological systems, or, to use a more correct term, living beings display just the kind of behaviour that is systematically different and, moreover, that runs just the contrary to the laws of physics and chemistry (Bauer, 1935). A living organism initiates processes that alter the direction of methods set up according to physical and chemical laws, and in just such a way that under the newly generated, different circumstances there will occur courses of action having just the opposite direction than they should take according to the laws of physics within the initial conditions. Instead of approaching equilibrium, the result of the biological processes will be the preservation and increase of the distance from the physical equilibrium. If the principle of life did not exist as a separate and independent principle from physics, then the accidental initiation of biological processes would, after a short period, quickly decline towards the state of equilibrium, towards physical 'equilibrium death', to generalise the concept of 'heat death' in order to include more than thermodynamic equilibrium. But as long as biological laws are irreducible to physical ones, the tendency towards physical equilibrium due to the balancing tendencies of the different physical or chemical gradients cannot prevail, for they are overruled by the impulses arising from the principle of life. Of course, these impulses arising from the principle of life are also material forces, as they manifest themselves in natural processes, and they also elicit natural processes. The main point is that biological impulses have a nature that elicits, maintains, organizes and coalesces the processes, which may otherwise set up only stochastically, transiently, unsystematically, and incoherently when physical principles are exclusive. The essential novelty of the biological phenomenon therefore consists in following a different principle, which is able to govern the biological phenomena even when the physical principles keep their universal validity. Until a process leads to a result that is highly improbable according to the laws of physics, it may be still a physical process. But when many such extremely improbable processes are elicited, and they are co-ordinated in a way that together they follow a different principle thereby making these processes a stable, long lifetime process, then we encounter a substantial novelty, which cannot be reduced to a lower level principle.

An analogy may serve to shed light on the way in which biology acts as compared to physics. It is like Aikido: while preserving the will of the attacker and modifying it using only the least possible energy, we get a result that is directly opposed to the will of the attacker. It is clear that the ever-conspicuous difference between living beings and seemingly inanimate entities lies in the ability of the former to be spontaneously active, to alter their inner physical conditions according to a higher organising principle in such a way that the physical laws will launch processes in them with an opposite direction to that of the 'death direction' of the equilibrium which is valid for physical systems. This is the Aikido principle of life. A fighter practising the art of Aikido does not strive to defend himself by raw physical force. Instead, he uses his skill and intelligence to add a small power impulse, from the right position, to the impetus of his opponent's attack, thus making the impetus of the attacker miss its mark. Instead of using his strength in trying to stop a hand coming at him, he makes its motion faster by applying some little technique: he pulls on it. Thus, applying a little force, he is able to suddenly upset the balance of the attack, to change it, and, with this, to create an advantageous situation for himself. The Aikido principle of life is similar to the art of yachting. There, too, great changes can be achieved by investing in small forces. As the yachtsman, standing at the helm of his little ship, makes a minute move to shift his weight from one foot to the other, the ship sensitively changes its course. Shifting one's weight requires little energy, yet its effect is amplified by the shift occurring in the balance of the hull. Control is not exerted directly on the surface level, but on the level of balance; it is achieved through altering this balance in a favourable direction against much larger forces. In this way, the effect of a very small force prevails. However, being able to alter this balance in a favourable direction presupposes a profound knowledge of contributing factors, as well as the attitude and ability to rise above direct physical relations, and the ability independently to bring about the desired change. If life is capable of maintaining another 'equilibrium of life', by a process the direction of which is contrary to physical equilibrium, then the precondition of life is the ability to survey, to analyze, and to spontaneously, independently and appropriately control all the relevant physical and biological states. Thus, indeed, life cannot be traced back to the general effect of the 'death magnet' and mere blind chance that are the organizational factors available for physics. The principle of life has to be acknowledged as an ultimate principle which is at least as important as the basic physical principle, and which involves just the same degree of 'objectivity' as the physical principle. If it is a basic feature of life that is capable of displaying Aikido-like effects, then life has to be essentially different from the inanimateness of physics, just as the principle of the behaviour of the self-defending Aikido disciple is different from the attacker's one. Thus, in the relationship of the laws of life and those of physics, two different parties are engaged in combat, and the domain of phenomena of two essentially different basic principles are connected. Practising the art of Aikido is possible only when someone recognizes and learns the principle and practice of Aikido. Now, regarding the origin of the principle of Aikido, it results from the study of the art of fighting. Regarding the origin of the principle of biology, it cannot result from the physical laws by a corporeal principle, since the ultimate principle of physics acts in contradiction to the life principle. Therefore, the life principle appears to be an ultimate principle behind the realm of physics.

It is an essential constituent of life that events that are extremely improbable on a purely physical basis do not occur in a sporadic or haphazard way. Instead, the occurrence of such events is the characteristic feature of life, being systematic and recurrent. Processes contrary to the laws of physics do not occur only once in a while during billions of years, but they are taking place abundantly and continually, and, in addition to that, such processes themselves provide the conditions for their own reproduction. The appearance of life is like a self-supporting avalanche: the influence that is exerted upon the system is amplified over and over again. In living systems, the direction of the growth of organization also differs from that in lifeless systems. According to the physical laws, entropy, or the measure of the disorder of lifeless systems, either constantly increases or remains at the same level. But in living systems, the level of organisation grows in the process of individual and species evolution. The organization of mammals, for example, is greater than that of insects. At the same time, the nutrition and the metabolic processes of the organism do not necessarily entail such an accompanying decrease in the organizational level of the environment that can otherwise be expected on the basis of physics. Still, the world does not divide itself into well-separated living systems and inanimate environment, with ever growing chasms between them, with the entropy of the one side always increasing and that of the other side always decreasing. And this points to the fact that the whole of the living kingdom and the biosphere can be conceived of and treated as alive.

3.1 *The Sun as a living being*

Recent astrophysical research has raised the possibility that the Sun (see Grandpierre, 1995a, b, 1996a, b, 1999a, 2000a) and the whole of the Solar System might enjoy a living nature. The Sun shows the phenomenon of self-inducing activity characteristic to living beings in its solar activity. Nuclear resources act to extend the lifetime of the star-state of the Sun by a factor of a million. Therefore nuclear burning seems to play a role for the Sun similar to what nutrition plays for earthly organisms. Moreover, the Sun is shown to be in an ultra-sensitive state in its basic activity phenomena. The astronomically very small effects of planetary tides (and perhaps planetary electromagnetism) on the Sun may play an important role in initiating stellar explosions in the solar core when the tidal waves interact with magnetic fields. This interaction produces locally macroscopic electromagnetic heating, which in the solar core may lead to a positive feedback since the rate of nuclear reactions grows in a high power of temperature. Local warming leads to the acceleration of nuclear reactions, which, in turn, produce more heat, which accelerates again the nuclear reactions - and so hot bubbles are generated in an explosive manner, which will be accelerated to the outer layers where they may produce atmospheric explosions known as solar flares. This ultra-sensitivity of the solar activity is basically similar to the stimulus-response process, which is regarded as a basic life criterion of earthly beings. This is even more the case, since not only the outer influences of planets, but the local free energy content of the Sun are what determines which 'stimulus' is to be amplified and how. These new results posit the question of the nature of the Sun within a novel perspective, one which is readily accessible to scientific inquiry. Following this train of thought, a brain-approach to understand the connections between Man and the Universe has been recently developed (Grandpierre, 1999c). Thus the physical point of view in the study of nature seems to be valid only in the artificial and abstract system of the most restricted and immediate conditions.

3.2 *The Life Principle*

In our contemporary world, natural philosophy, which once used to comprehend the whole of Nature, is co-extensive with materialism. Materialism seeks an explanation of the world in corporeality, in 'objective entities'. In physics, this appears as the atomising attitude, the seeking for the final constituent parts of Nature, and its elementary particles, and trying to understand the world from them. But how can the world be understood on the basis of an atom? This question makes it necessary to consider the relationship between material entities and a non-material basic principle. To be able to get a deeper understanding of this object-law relationship, let us first examine the fundamental principle of biology, the life principle.

      The founder of exact theoretical biology, Erwin Bauer has proven in his fundamental work *Theoretical Biology* that the basic laws of biology cannot be traced back to the laws of physics (Bauer 1935, pp.24-9). Erwin Schrödinger's classic study *What Is Life?* (1944) led physicists to biology. Every biological system (although this expression itself is a compromise with polymorphic materialism, for 'system' suggests lifelessness), that is, every living being is a living being precisely because it displays independent and spontaneous changes, changes that cannot be traced back to the physical conditions prevailing under the given circumstances or to the equations of physics. Bauer gives us the following definition of the life principle in his work: 'Living systems and only living systems are never in a state of equilibrium; to the debit of their reserves of free energy they continually perform work against the setting in of the state of equilibrium which should, according to the laws of physics and chemistry, be established under the actual external conditions'. So the key issue is equilibrium. It is clear that a

living system can only reach a total, temporally stable, static physical equilibrium after its death, since as long as it is alive, it is capable of producing spontaneously active processes. Yet there is a widespread notion that living beings are in a dynamic equilibrium. What does this dynamic equilibrium mean? It means that although the system is stable and it does not change in time, its stability is maintained by the equilibrium of different processes – just as a waterfall will always looking the same, and constant in time. This is because water keeps falling down steadily. Stability can only be maintained by allowing the same amount of water to flow in as the amount that flows out. We know that living beings display stability, and we also know that there are changes inside them at all times – so in that respect, humans and waterfalls seem to be alike. Moreover, humans are capable of performing work, and so are waterfalls. But does this means that we humans may be regarded in this respect as similar to waterfalls? Is there the same amount of material flowing in as out, and is it this that maintains our stability? Bauer has shown that this similarity is only superficial, and behind it there lies an essential difference. The stability of the waterfall depends on the water supply. As the amount of inflow decreases, the amount of outflow decreases in just the same way and pace; there is a clear and simple relationship between the two, expressed by the equality of the two quantities. If there were no water supply, there would be no waterfall. Living systems are essentially different from inanimate systems, because if the food supply of a human is stopped, her/his life would not disappear. Humans are not food-flows. The essential difference means that living beings are capable of spontaneous activity, and this spontaneous activity, the capability of living systems to perform work, does not originate from outside, but from inside. A waterfall is unable to modify its own conditions, but being able to do that is precisely the characteristic feature of the living being. If a human is left completely without a supply of food he would die only after a period much larger than the time needed for the food to flow through the body. Moreover, it is characteristic of living beings that they act against their unfavourable circumstances. They are able to produce from themselves such new organisms (their offspring) which themselves are capable of producing further living beings – thus the living organism lives on in another form; its life is sustainable continuously for cosmic ages. It is obvious that the human organism's internal source of energy, its free energy (i.e. its ability to perform work) is finite. So it cannot be expected of the human organism that it should, after that free energy runs out, perform any spontaneous activity that needs any kind of a source of power.

The question is precisely this: whether a given system or being has any free energy, and if it has any, does it use it for processes directed towards the physical equilibrium, or, on the contrary, to increase its distance further from the physical equilibrium? Free energy, or the ability to perform work, can be mobilized from the inside by the living organism itself. The sources of energy that waterfalls (or flames) need to maintain their ability to perform work (to maintain their distance from the state of physical equilibrium) can be found outside of the system, whereas in living systems these sources of energy can be found inside it. Living organisms are able to ensure, on their own, the maintenance of their ability to perform work. Applied to a waterfall this would mean that the waterfall itself provides for its water supplies, carrying the already fallen water back up on its own, so that it can fall down again and again. It has to be pointed out that if the energy that is needed to lift the water up is acquired from the water's kinetic energy by transforming it back, that still does not mean that the waterfall is a living system. In such a case, the energy of the already fallen water (which is now outside the waterfall) is transformed such as happens in a hydroelectric power plant, where the kinetic energy of the water produces electricity. That electricity can then be used to power a pump, which lifts the water up even though the net efficiency can never be 100 per cent. Still this appliance is outside of the waterfall.

In living systems, the corresponding appliance is to be found inside the living beings, under the complete supremacy of that given living being, so it can be mobilised at any time. To be able to do away with the ideal difference between a waterfall and a living organism, the waterfall should have to be able to produce and maintain an otherwise continuously shrinking level difference at a constant rate which is needed to make water fall down. Let us imagine a waterfall that produces its level difference itself: all of a sudden, there starts to appear a volume differential in a river, which then grows, stopping after some time at a certain point. This differential would not be produced by some earthquake, but by a delicate, intrinsic and almost ethereal process, when we want to preserve the analogy with living beings.

The necessary transforming appliances that would be able to produce the difference in the level of a waterfall can essentially be found in us. However, living beings find themselves facing various external conditions moment by moment, to which they have to react differently to be able to maintain their capacity to live. In our parallel above, there is an analogy between the free energy of the living organism and electricity. But as the given situations and the kinds of work that the living organism has to perform are different and varied, in one situation the living organism will need a pump, in another, a wing, in yet another, a spring. In almost every situation a different kind of machine is needed. This means that the creature has to produce different kinds of machines continuously – and thus, an infinite row of different pieces of equipment. And all of this creation of machines should be done in an orderly manner. Living organisms do not work only under the influence of momentary impressions; instead, they also follow long-term organizing principles – that is, they 'work rationally'. This 'reasonable' activity is another mystery, to which we shall have to return. But the requirement of producing devices continuously, almost moment by moment, and in accordance with the given situation, is inevitable, since we have to support and maintain the activity of our organism, the 'outer work' of our body, and we have to exert real physical effects to reach these aims. Instead of mechanical pumps, we produce such microscopic appliances in our organism that can generate the movement of our body and muscles. These tiny machines that control the muscle system have to be capable of mechanical power impulses, of bringing about contractions in muscles, and at the same time they have to do this from one moment to another and on the basis of reason.

It is interesting and also worthwhile to consider to what extent we human beings are taking part in our own vital functions. To what extent are we, as conscious beings, participating in the activity of our organism, of our intrinsic organization, of our vital force? Are we doing our best to distance ourselves as far as possible from morbidity? Are we using all our powers to increase our capacity to live? Are we directing all our powers against those physical and chemical forces that affect us in a given situation? Do our lives move in the line of least resistance, like that of a stone thrown away? Can we consciously feel the attraction of our intrinsic vital force, and do we carry it on, reinforcing it by living a consciously led life? Do we really shape our lives in an independent and spontaneously active way, in accordance with the attraction of the ultimate principles of our lives? We, as living beings, should thoroughly consider what possibilities we are given by revealing the ultimate nature of our lives, our life principle.

Now that we have obtained a picture of the life principle, let us examine how the life principle may exert its influence. How can the constituent parts of a living organism recognize and follow this life principle? Do they perhaps perceive it as we humans do? They perceive it in a different way than the way we perceive things, since between our perception and our behaviour there are many intermediate levels, among which – as shown by our inquiry – one of the essential ones is the regulating, controlling, amplifying, directing and guiding effect of the life principle. Thus, for beings without self-consciousness, perceiving the life principle does not include a process that is under the control of another development. Because, for beings without self-consciousness, this route is without any transmission, the perception and expression of the life principle are made not indirectly but directly, and this suggests that the one that perceives and the one that is expressed are almost identical. For us, transmissive beings, who have not only one, but three ultimate driving principles with transmissions between them, this near identity may seem strange, sometimes even mechanical or automatic. We know that machines work exclusively on the basis of the physical principle, that is, without the transmission of another principle. Now we may return to deal with the question: how are elementary particles capable of perceiving the fundamental physical principle?

3.3 *Matter or principle? Is materialism physical or spiritual?*

The fundamental principle of materialism and 'physicalism', or the physical world concept, is the principle of inert matter. The science of physics derives all its results due to this 'inert material principle'. This principle made it possible to recognise the real connections between seemingly discrete phenomena, and this principle made it possible to represent and describe the physical phenomena in a mathematical and logical form. There is a significant difference between talking of the atom with the knowledge of the laws and principles of physics and, generally, with logic (Grandpierre 2000a, pp. 19-20) in mind, on the one hand, and talking of the atom without such knowledge in mind on the other. Without a comprehensive and inevitably spiritually organizing factor, physics would have no laws, nor would logic – and thus we would simply be unable to conceive of something as sterile and abstract as the notion of the atom. Without the factor of an organizing principle, materialistic nuclear physics would never get further than the sterile concept of the atom. Materialism, being 'matter-principled', is founded on corporeal/material and inevitably on principal/spiritual grounds at the same time. Materialism builds up on a material and a spiritual factor: on atoms and on physical laws. There are no atoms without physical laws, and this fundamental fact shows that there is no materialism without spiritualism. Sheer materiality, the concept suggested by materialism, is essentially a self-contradicting concept, denying its own generating factor, the principle behind which the concept generates itself. It is easy to see that a sheer materialistic view, one that would really deny all spirituality, would be like the perfect embodiment of closedness and readiness: it would be an eternally inert, inanimate world without laws and understanding. Therefore, we have to keep in mind that actually materialism essentially represents a kind of obscure spirituality, but that in fact this spirituality has declared the denial of its own spiritual nature as its basis.

3.4 *Are interactions material?*

Important conclusions can be drawn here. According to the materialistic view, apples fall on the ground because the Earth attracts them. But how is the Earth able to exert attractive force? By what device? Does it emit an attractive effect? What is the nature of that effect – is it material or spiritual? How can the attractive effect exert this attraction? Does it let out some kind of matter from itself, for example gravitons in the case of gravitation? But if this would be so, the amount of gravitation of bodies should decrease over time. Similarly, electric charges could not remain strictly static at a constant charge, because they would have to emit electro-magnetic energy permanently, and thus the force field of the charges would have to diminish. Even if one would acknowledge the fact of energy and mass emission from gravitational, and electrical charges, one could speculate that they could get back the same amount of energy-mass from the quantum-vacuum field of the universe. Even in this case it would be necessary that a principle should exist which would continuously regulate all the charges of the universe so that the energy-mass exchange input and output would be balanced. In this case, we again would reach a picture in which a principle regulates the mass and energy flows of the universe. But how can a spiritual principle exert a material effect? Another example would be the fact that contemporary science regards the value of electric charges as a universal and unchangeable constant, although there are some theories about gravitation diminishing over time. If we disregard the hypothetical universal balancing mechanism, then we have no other choice but to admit that the influences, be they electro-magnetic or gravitational, that bodies emit are not of a material nature, since all matter has energy and a corresponding amount of mass. Now since a nascent charge has a material influence, it should emit a material influence, therefore produce energy from itself, and so its charge should decrease, which is not the case. From this we can deduce that we are dealing here with effects that can be described with mathematical exactness, but which are not exclusively material effects. In both of these cases, the question arises: where does the ability of matter to exert influences come from if its material substance remains constant?

3.5 *Clever matter?*

How do atoms know the laws of physics? How does the wind know which way to blow under any given circumstance? It knows this because the power arising from the differential pressure drives it towards areas where the pressure is lower. But why does matter migrate to places with a lower pressure? The standard response is because the laws of physics prescribe this. Eventually, this comes down to the principle of least action. But then again, how can a principle cause a physical effect? How can a spiritual factor be able to move matter? And the ultimate question of physics is why there are there physical laws at all. How can any body follow the principle of least action? The answer to this question is similar to the explanation given to the path that light follows. Light travels between two points along the shortest possible route, even if there is a mirror in its way somewhere along its course. How is light able to select the shortest route? When Feynman introduced the path-integral principle, he pointed out that to be able to follow the principle of least action, light (or any other quantum process) must 'virtually' go over all the possible routes, over all the possible histories, and then these add up to the 'actual' shortest route. The precondition of such an adding up is that in the course of surveying the possible routes, light virtually has to travel over all the routes at a speed much larger than the velocity of light, so that by the time it comes to the adding up, the travelling speed of light on the actual route should be equal to the velocity of light. Feynman has put all this into a mathematical formula – but how is it possible that a lifeless and sterile atom can do all that? How can a perfectly abstract atom perceive a principle and behave according to it? Is there such a spiritual factor that is capable of exerting physical effects?

These questions raise the problem of the origins of physical laws. In the contemporary physical world concept, apparently this problem cannot be accounted for on a scientific basis. But excluding the question of the origin of the physical laws from the scope of science is a refutation of the original aim of science, namely to understand nature. Science cannot declare that it is a scientific taboo to examine the laws of those levels of Nature that are deeper than the physical level. If present-day science does so, we can be sure that that is an unscientific and anti-cognitive attitude.

# 4. HOW CAN WE EXAMINE THE PROBLEM OF THE ORIGIN OF PHYSICAL LAWS WHILE PRESERVING THE AIM OF SCIENCE: TO UNDERSTAND NATURE THROUGH VALID AND SYSTEMATIC KNOWLEDGE?

*4.1 The origin of physical laws*

From where do the laws of physics arise? Sir Arthur Stanley Eddington (1882-1944) the Cambridge University pioneer in cosmology, suggested in the frame of the present-day scientific views that the maintenance of these laws is not supported by any factor behind these laws. 'He believed that a great part of physics simply reflected the interpretation the scientists imposed on his data' (Vibert 1979, Vol. 6, p. 298). Therefore these laws are law-like only as an ultimately improbable chance event. The laws can be present from the beginning of the Universe only by a chance coincidence of events - it is not possible to exclude the case when from the completely random collisions of atoms an illusion develops. It has to be clear, says Eddington, that from the fact that a stone falls down after a hundred throws does not necessitate with an absolute necessity that the stone will fall dawn after the $101^{st}$ time one throws it up. And if it falls down at the end, it is only a mere result of chance again. Therefore, there are no physical laws in the Universe: the apparent lawfulness is a result of an extremely rare ultimate coincidence of random events.

I do not think that we can be satisfied with such a description, which does not reach the causes and remains in the realm of phenomena only. The term chance expresses in that context only that the cause of the phenomenon studied is not known. Therefore, chance cannot explain any phenomena since 'explain' implies setting up a relation, which explains the yet unknown with a known. So we do not think that the interpretation of the origin of physical laws as being the result of a mere chance would explain anything. Now the basic physical principle, the principle of least action, is not itself a phenomenon. On the contrary, emphasizing the difference between the absolute validity of physical laws and the randomness of human activity suggests that the automatic mechanistic character of the prevalence of the physical laws implies that atoms could not have free will at all. But is it really true that in the atomic world every event goes ahead with an absolute necessity? We are far away from thinking it true in the realm of physics itself. Not only chaotic phenomena, but also more evidently the phenomenon of life is developed with the assistance of atoms. Therefore if life grows out of the atomic world, then the atomic world would have to be able to generate the biological laws which are able to govern the physical laws within the conditions present in living organisms, which are also generated by the atomic world. Now we have to see that although it would not seem necessary for a physicist, since it does not follow from the laws of physics, the atomic world has to be able to make free decisions, and it realizes such free decisions during its activity when the proper conditions are reached. The decision of the atomic world is in favour of the development of conditions necessary to life itself. And now we can consider the question: is it true that strictly consequent behaviour is possible only by machines?

To be able to answer this question it is necessary to find the proper context. This is needed since it is easy to state that atoms are not alive, since atoms cannot speak to us, nor do they have journals to officially refute the claim of their inanimateness. Moreover we know that Descartes maintained that not only atoms but animals are also inanimate. But we may realise now that there is an essential difference between machines and living organisms, namely that organisms have inner motivations, which may dominate over the outer conditions, while machines do not have such. It is true that it is not easy to demonstrate from an inner motivation that it is truly inner. But the question is settled by the theoretical biology of Ervin Bauer who recognized that living beings follow a deeper organising principle than the inanimate objects. Therefore it is now possible to draw proper conjectures from the fact that the realm of animals is not so far away from us.

4.2 *Atomic instincts*

Apparently, in human consciousness the governing factor is awareness or self-consciousness. The self-conscious reason is free in the sense if it is able to decide within certain conditions according to its own viewpoints. At the same time, self-consciousness is based on a consciousness-without-self-reflection. This consciousness without self-reflexivity, as nature shows us, possesses reason and this is why animals act reasonably, grow and act coherently according to their needs. Self-consciousness is only a late offspring of the reasonable natural consciousness. So our actions are not only governed by self-consciousness, but also by this natural consciousness. We see in the animal kingdom that the behaviour of animals is largely governed by instincts, which are such universal, almost inevitable motives moving the animals. As compared to humans, it is apparent that the behaviour of animals is governed to a larger extent and degree by instincts than the functions of humans. We may recognise that the effect of instincts in human behaviour is not realised with such a strict necessity as that of animals. In fact, we may find that instincts are more formal in more primitive animals, and they may be even more so for plants. When continuing our considerations up to the atomic world, the progressive rigidity of instincts seems to be continuing at the atomic world as well. It is reasonable to assume that the instincts are more formal in the atomic world, than in the kingdom of plants, animals and humans. This is apparently only a difference in degree, not in substance. Therefore, it seems to be plausible to conjecture that instincts should be the strongest in the atomic world.

We hypothesize that atoms are less independent from their own inner drives than humans are. This conjecture may also be evaluated in that atoms show more deterministic behaviour than people. Being more deterministic, it is easier to describe their behaviour with mathematics. But this does not mean necessarily that they do not think and are not able to decide freely within certain conditions proper to them, since they have already proven their freedom by developing life and consciousness where they are able to behave more freely. Doing this, they have transcended the laws of physics and have developed the realms of biology and psychology. The fact that it is possible to describe a phenomenon mathematically does not imply that the entity that realizes the phenomenon is dead. For example, it is possible to describe the motion of an animal rolling down a hill in mathematical terms, but this does not imply that the animal is dead.

Therefore, from the fact that atoms are not as free as people, it is not valid to conjecture that atoms are inanimate objects. On the contrary, we can observe that the instincts are demonstrated the most directly and naturally in the behaviour of atoms, especially in a long baseline timeframe. This conjecture opens up a new perspective for the study of the general instincts of nature. We would have obtained a solution to a riddle unsolved for thousand years. Both science and religion have declared the question of the origin of physical laws as insoluble or inconceivable. Now we have offered an important insight to the nature of this riddle: physical laws develop as a consequence of phenomena very similar to that of the human world, namely inner motivations, inner principles are their ultimate movers. Therefore, the origin of physical laws leads us to the inner world of nature, untouched until now. This is the realm from where the laws of physics arise. In this way we have obtained a context of interpretation, which, instead of separating physics from philosophy, connects them together, and doing this it expands the usual range of scientific investigation.

One may observe that materialism, when attempting to remove all spirituality from the universe, cuts itself off from being able to understand the origin of physical laws. Atoms are able to follow the laws of physics because they perceive the world of their own instincts. Their bonds to their world are more rigid than that of humans to theirs. Expressing the difference with a metaphor, atoms are small skiffs driven by the flow of the river of their instincts. Man, in comparison, appears more like a sailing-boat, and bonded also to another principle, he is able to navigate with both drives: the drive of the flow below, and the drive of the wind above. This is why Man is able to proceed less rigidly and faster. But the wind is also the result of the free energy, the heat surplus of the atoms of the river. And the atoms of the river also create the sail of the larger boats by their own decisions in this metaphor.

Our considerations suggest that the hypothesis of identical elements of particle physics have to be tested experimentally, suggesting concrete experiments which could explore for example how the individual history of large amounts of radioactive materials do influence the radioactive decay of a radioactive element under intensely regular laser impulses with strict frequencies occurring in strict periods (E. K. Grandpierre, 1991, 2000).

Our result is that the considerations of the origin of physical laws do not lead us away from the domain of science. On the contrary, our search for the origin of physical laws have supplied us with scientific evidence that the range of science extends beyond strict and exclusive materialism, involving not only the sovereign worlds of biology and psychology, but reaching as well into philosophy which has to be able to consider the ultimate meaning of the ultimate principles, the *archi*. Recognising the nature of these *archi*, we more or less living beings and more or less humans will become able to realise our own deepest nature with clearer and brighter prominence.

## 5. CONCLUSION

Our research has found deep and meaningful connections between the basic principle of physics and the ultimate principles of the universe: matter, life and reason. Therefore, the principle of least action is not necessarily an expression of sterile inanimateness. On the contrary, the principle of physics is related to the life principle of the universe, to the world of instincts behind the atomic world, in which the principles of physics, biology, and psychology arise from the same ultimate principle. Our research sheds new light to the sciences of physics, biology, and psychology in close relation to the basic principles. These ultimate principles have a primary importance in our understanding of the nature of Man and the Universe, together with the relations between Man and Nature, Man and the Universe. The results offer new foundations for our understanding our own role in the Earth, in the Nature and in the Universe. Even the apparently inanimate world of physics shows itself to be animate on long timescales and having a kind of pre-human consciousness in its basic organisation. This hypothesis offers a way to understand when and how the biological laws may direct physical laws, and, moreover, offers a new perspective to study and understand under which conditions can self-consciousness govern the laws of biology and physics. This point of view offers living beings and humans the possibility of strengthening our natural identity, and recognising the wide perspective arising from having access to the deepest ranges of our own human resources and realising the task for which human and individual life has been created.

**Acknowledgement**

The author wishes to express his thanks to Dávid Galántai in preparing the translation of this text.

# REFERENCES:

Aspect, A. and Grangier, P. 1987. *Hyperfine Interactions* 37.3.

Atkinson, P. W. 1981, *Creation*, W. H. Freeman and Co., Oxford.

Bauer, E. 1920. *Die Grunprincipien der rein naturwissenshtlichen Biologie*. Berlin.

-         1935 (1967). *Theoretical Biology*. Moscow-Leningrad: IEM.

Bunge, M. 1967, *Scientific Research*, Springer-Verlag, Berlin Heidelberg New York, Vol. 1, 291.

Delsemme, A. 1998, *Our cosmic origins. From the Big Bang to the emergence of life and intelligence*. Cambridge Univ. Press.

Diogenes Laertius, 1959, *Lives of eminent philosophers,* Harvard Univ. Press.

Grandpierre, A. 1995a, *Quantum-Vacuum Interactions in the Brain,* Appendix, in Ervin Laszlo: The Interconnected Universe. Conceptual Foundations of Transdisciplinary Unified Theory. World Scientific, 1995.

-          1995b, Peak Experiences and the Natural Universe, Invited Essay. World Futures. *The Journal of General Evolution*. 44: 1-13.

-          1996a. A Pulsating-Ejecting Solar Core Model and the Solar Neutrino Problem. *Astronomy and Astrophysics* 308: 199-214.

-          1996b. On the Origin of Solar Cycle Periodicity. *Astrophys. Space Sci*. 243: 393-400.

-              1999a. A dynamic solar core model: on activity-related changes of the neutrino fluxes. *Astron. Astrophys.* 348: 993-999.

-          1999b. A Nap életközpontja (The life centre of the Sun), *Harmadik Szem* (Third Eye), July-October.

-        1999c. The Nature of Man-Universe connections. *The Noetic Journal* 2: 52-66.

-        2000a. The nature of the universe and the ultimate organizational principle. *Ultimate Reality and Meaning*. 23: 12-35.

-        2000b. The thermonuclear instability of the solar core. *Nucl. Phys. B.* (Proc. Suppl.) 85: 52-7.

Grandpierre, K. E. 1991. personal communication.

-        2000. Collective Fields of Consciousness in the Golden Age. *World Futures. The Journal of General Evolution.* 55: 357-79.

Grangier, P., Roger, G. and Aspect, A. 1986. *Europhysics Letters*, 173.

Grangier, P. 1991. Wave-particle duality for the photon. From basic concepts to present experiments. *Zeiss Information* 31: 102.

Hawking, S. 1988, *The Brief History of Time from the Big Bang to Black Holes.* Bantam Books

Hoyle, F., Wickramasinghe, N. C. 1999a. The Universe and Life: Deductions from the Weak Anthropic Principle. *Astrophys. Space Sci.* 268: 89-102.

-        1999b. Biological Evolution. *Astrophys. Space Sci.* 268: 55-75.

Landau, L. D. and Lifshitz, E. M. 1959. *Course in Theoretical Physics*. Vol. I, translated by J.B. Sykes and W. H. Reid. London: Pergamon Press, §2, p. 12.

Quay, P. 1995. Final Causality in Contemporary Physics. *Ultimate Reality and Meaning* 18: 3-19.

Schrödinger, E. 1944, What is Life? Cambridge: University Press.

Vilbert, D. A. 1988, Eddington, Sir Arthur Stanley. The New Encyclopedia Britannica, 30 Volumes, Vol. 6, Chicago: Encyclopedia Britannica Inc., 298.